\documentclass[aip,rsi,twocolumn,showpacs,reprint,numerical,amssymb]{revtex4-1}
\usepackage{graphicx}
\usepackage{color} 
\usepackage{upgreek}
\usepackage{amsmath}

\definecolor{darkblue}{rgb}{0,0,0.5}
\definecolor{lila}{rgb}{0.3,0,0.3}
\definecolor{turq}{rgb}{0,0.1,0.4}
\usepackage[colorlinks=true,
linkcolor=darkblue, 
  filecolor=red,
  citecolor=turq, 
  urlcolor=lila 
]{hyperref} 

\newcommand{\degree}{\ensuremath{^\circ}}

\begin{document}

\title{A Rubidium M$_{\mathrm{x}}$-magnetometer for Measurements on Solid State Spins}
\author{Daniel Arnold}\affiliation{3. Physikalisches Institut and Stuttgart Research Center of Photonic Engineering (SCoPE), Universit\"at Stuttgart, Pfaffenwaldring 57, Stuttgart D-70569, Germany}
\author{Steven Siegel}\affiliation{3. Physikalisches Institut and Stuttgart Research Center of Photonic Engineering (SCoPE), Universit\"at Stuttgart, Pfaffenwaldring 57, Stuttgart D-70569, Germany}
\author{Emily Grisanti}\affiliation{3. Physikalisches Institut and Stuttgart Research Center of Photonic Engineering (SCoPE), Universit\"at Stuttgart, Pfaffenwaldring 57, Stuttgart D-70569, Germany}
\author{J\"org Wrachtrup}
\affiliation{3. Physikalisches Institut and Stuttgart Research Center of Photonic Engineering (SCoPE), Universit\"at Stuttgart, Pfaffenwaldring 57, Stuttgart D-70569, Germany}
\affiliation{Max Planck Institute for Solid State Research, Heisenbergstra\ss e 1, D-70569 Stuttgart, Germany}
\author{Ilja Gerhardt}
\affiliation{3. Physikalisches Institut and Stuttgart Research Center of Photonic Engineering (SCoPE), Universit\"at Stuttgart, Pfaffenwaldring 57, Stuttgart D-70569, Germany}
\affiliation{Max Planck Institute for Solid State Research, Heisenbergstra\ss e 1, D-70569 Stuttgart, Germany}
\date{\today}

\begin{abstract}
The detection of environmental magnetic fields is well established by optically pumped atomic magnetometers. Another focus of magnetometry can be the research on magnetic or spin-active solid-state samples. Here we introduce a simple and compact design of a rubidium-based M$_{\mathrm{x}}$-magnetometer, which allows for hosting solid-state samples. The optical, mechanical and electrical design is reported, as well as simple measurements which introduce the ground-state spin-relaxation time, the signal-to-noise ratio of a measurement, and subsequently the overall sensitivity of the magnetometer. The magnetometer is optimized for the most sensitive operation with respect to laser power, and magnetic field excitation at the Larmor frequency.
\end{abstract}

\maketitle

\section{\label{Introduction}Introduction}
To date, the detection of magnetic fields at various frequencies and length-scales is an indispensable tool in applied science. Material science~\cite{zhang_as_1995}, geological explorations~\cite{becker_ap_1995}, and the characterization of biomagnetic fields~\cite{bison_apb_2003} are performed with an accuracy, which was not available some decades ago. Most recently, the combination of magnetometers with nuclear magnetic resonance~\cite{ledbetter_jmr_2009} and its related imaging techniques are present in every-day applications. Besides these very technical approaches, a number of current research topics provides insight towards novel sensing applications. Another magnetometry related research field is the implementation of high precision tests in fundamental physics~\cite{weis_ole_2005}.

Early implementations developed from the detection of (macroscopic) electrical currents towards the detection of quantized entities e.g.\ flux quanta in a superconducting interference device (SQUID). These devices exhibit record sensitivities in the aT/$\sqrt{\mathrm{Hz}}$-range at low frequencies and are commercially available. The availability of liquid helium limits their applicability, especially in off-site applications. In the last two decades, the early experiments on atomic vapors~\cite{robinson_baps_1958} quickly revealed their sensitivity on external magnetic fields and led to optical pumped magnetometers~\cite{budker_np_2007}. Current devices reach sensitivities comparable to the mentioned SQUIDS~\cite{kominis_nature_2003}, but have relaxed requirements in terms of cryogenics.

In parallel to the field of optically pumped atomic magnetometers, the research on spin active defects in diamond has emerged. The best researched defect center is the negatively charged nitrogen-vacancy center~\cite{gruber_s_1997}. It can be conveniently optically prepared and enables the sensing of environmental spins and the implementation of quantum optical primitives. Early experiments on their coupling to other quantum devices have been performed~\cite{kubo_prl_2010}.

Optically pumped magnetometers based on atoms can be implemented in a variety of experimental schemes~\cite{budker_book_2013}. In any case, an (external) magnetic field changes the influence on a near-atom-resonant probe beam. Two fundamental schemes are well established: The M$_{\mathrm{z}}$-magnetometer, which detects the fingerprint of the Larmor precession of an atomic ensemble by a steady change in intensity on a photo detector. The other scheme is the M$_{\mathrm{x}}$-magnetometer. It directly measures the Larmor frequency on the optical signal, which is commonly introduced by a 90\degree -configuration, comprising a pump and a probe beam. Many derived schemes are established in research, often focusing on special frequency ranges, such as SERF (Spin Exchange Relaxation Free) magnetometry~\cite{kominis_nature_2003} or the implementation of differential measurement schemes, such as the use of balanced detection, which suppresses the common mode laser noise. 

In this paper we present the implementation of a simple M$_{\mathrm{x}}$-magnetometer based on atomic rubidium. The design is adapted from the Fribourg atomic physics group~\cite{groeger_epj_2006}. The magnetometer design introduced in this work combines the pump and the probe beam, which makes it easier to handle. Since the future goal is a combination of this magnetometer and a solid-state spin ensemble, the device allows for hosting an experiment with NV-centers. Therefore, the specially tailored cell-design is described and the important parameters such as spin-lifetime and measurements for describing the magnetometry are introduced. And optimization procedure and the resulting sensitivity are presented at the end of the paper.

\section{\label{Experimental Setup}Experimental Setup}
\begin{figure}[!hbt]
\includegraphics[width=\columnwidth]{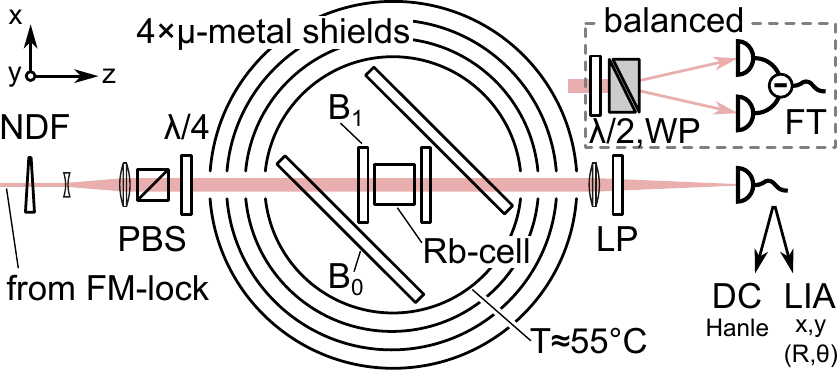}
\caption{Experimental setup of the 45\degree -M$_{\mathrm{x}}$-magnetometer. A locked laser beam passes a continuously variable neutral density filter (NDF) in order to control the intensity. The beam diameter is increased to $2\times w_{0}=$ 6~mm by a concave-convex telescope. The B$_{0}$ coils are aligned by 45\degree \ with respect to the optical axis, the B$_{1}$ coils produce a magnetic field co-linear with the optical path. The \mbox{(pseudo-)Helmholtz} pairs for the compensation of residual fields are not shown. The detectors are placed behind a long-pass filter (LP) and consist either by a normal amplified photo-diode or by balanced detection behind a Wollaston-prism (WP). Signal processing is either performed by a measurement card (DC) or a lock-in amplifier (LIA). The balanced photo-diode signal is processed by a Fourier transformation (FT).}
\label{fig:fig01}       
\end{figure}

The design criteria for the setup was having an easy implementable design of a rubidium magnetometer with a reproducible sensitivity. One of the advantages of the M$_{\mathrm{x}}$ design is the possibility to allow the detection completely based on alternating currents (AC). This suppresses a number of unwanted noise effects, such as laser power fluctuations, and other sources of $1/f$, and $1/f^2$-noise.

The common way of implementing an M$_{\mathrm{x}}$-magnetometer is a configuration with a pump and a probe beam. Another simplified way is to combine both beams and apply the reference B$_{0}$-field in an angle of 45\degree \ with respect to a single laser beam.

The optical setup of our M$_{\mathrm{x}}$-magnetometer is shown in Fig.~\ref{fig:fig01}. The laser source consists of a commercial Littrow system, operating at approximately 780~nm (DL Pro, Toptica GmbH, Germany). The fiber output of the laser is coupled out and split on a polarizing beam splitter. One part is fed into a Doppler-free frequency modulation rubidium spectroscopy. This allows for a spectral lock, which is performed by a commercial locking system (Digilock, Toptica GmbH, Germany). For the presented experiments, the laser is locked to the D$_2$-transition (5$^2$S$_{1/2}$ $\to$ 5$^2$P$_{3/2}$, F=3 $\to$ F'=2). The other part goes to the magnetometry setup. The power can be changed with a neutral density filter wheel, which is calibrated against the input power into the magnetometer. Then the beam is guided upwards to the level of the magnetometry setup (beam height 350~mm) at which point a mirror directs it horizontally towards the $\upmu$-metal chambers. On a small platform the beam is widened to approx.\ 6~mm beam diameter by a telescope. Before entering the measurement setup, the polarization is once more aligned by a polarizing beam splitter and a quarter wave plate to $\sigma^+$-light to pump the atoms in a defined spin polarized state. Behind the magnetometry setup, the continuous beam is focused either on an amplified photo diode (PDA-36A/EC, Thorlabs, US), or alternatively on a balanced detector (Nirvana 2007, Newport, US). To suppress unwanted light, we have introduced a 750~nm long-pass filter, such that even with the lights in the laboratory, no characteristic 50~Hz line frequency is observed by the detector.

The light levels were kept at a minimum, to reduce power broadening effects. A range of the input power from approx.\ 1 to 600~$\upmu$W was measured in front of the magnetometer. The gain of the photo diode was set to either 30 or 60~dB, depending on the measurement performed.

The signal of the amplified photo diode is analyzed on either a computer based digitizing card (PXI 4462, National Instruments, Austin, TX, US) or fed into a lock-in amplifier (SR 830, Stanford Research, Stanford, CA). The lock-in output (X,Y, or alternatively R, $\Theta$) is again fed into the same measurement card (PXI 4462, National Instruments, Austin, TX, US) as before.

Parallel to this acquisition, the external magnetic fields have to be controlled. This is electrically realized by an analog output card (PXI 6361, National Instruments, Austin, TX, US), which directly drives the B$_{0}$, and B$_{1}$-field. Initial configurations to drive a current in the coils via a dedicated current driver were not required or did not allow for a fine enough resolution. The compensation fields are controlled by a precision power supply (MP2030, Hameg, DE). We use its voltage output in the mV range and verified the validity of Ohm's law by a 6 1/2 digits digital multi meter (PXI 4070, National Instruments, Austin, TX, US).

All cards are locked onto the same PXI internal clock system (10~MHz), such that all samples are concerted with $\upmu$s accuracy. All generated and all acquired data are locked to the correct phase of e.g.\ the alternating B$_{1}$-field.

A self-implemented \textit{Python}-software gives access to a fully remote-controlled measuring operation. 
The controlled parameters include the laser power, the lock-in amplifier, the photo-diode gain, the compensation coils' power supply, and the heater's temperature. Also, all out- and inputs of the PXI cards are remotely accessible, which are use for data acquisition as well to drive the B$_0$- and B$_1$-coils. The acquisition program allows for automated measurements based on markup-language based measurement files (YAML). Generally, the photo-diode and/or the lock-in signal is acquired while another parameter is scanned. It is not possible to continuously change the frequency when the lock-in is used due to limitations of the lock-in's internal phase-locked-loop clock to follow an external reference clock source.
 
\begin{figure}[!hbt]
\includegraphics[width=\columnwidth]{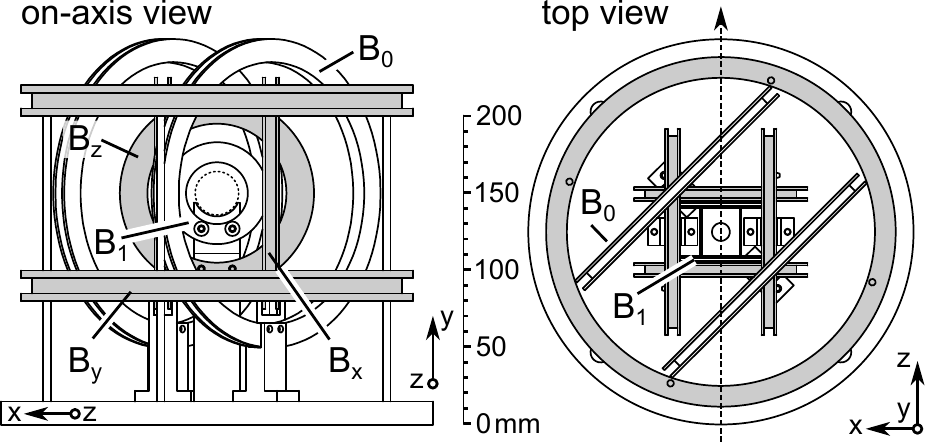}
\caption{\mbox{(Pseudo-)Helmholtz} coil configuration for the magnetometer. The coils for the B$_{0}$ and B$_{1}$-field are indicated in white. Compensation coils (B$_x$, B$_y$, B$_z$) are indicated in gray. The arrow indicates the optical path. From the bottom (through the holder for the atomic vapor cell) an optical fiber allows for an optical access to a solid-state spin ensemble.}
\label{fig:fig02}       
\end{figure}

The coils are designed to be able to set a static magnetic field in 45\degree \hspace{0.01mm} against the optical axis. Additionally, a B$_{1}$-field configuration is collinearly oriented to apply a radio frequency field to the atoms. The design principle is very comparable to earlier works. 

For a field compensation, another set of three \mbox{(pseudo-)Helmholtz} coils is introduced. These are shown in Fig.~\ref{fig:fig02}. For a complete description of the coil configuration, refer to Tab.~\ref{tab:tab01}.

\begin{table*}[htb]
\begin{tabular*}{\linewidth}{@{\extracolsep{\fill}}*8c@{}} \hline
\label{tab:compensation_coils_specs}
coil & R (mm) & d (mm) & N$_{\text{win}}$ & R$_{\Omega}$ ($\mathrm{\Omega}$)& B (mT/A) & $\Delta$B$_\bot$ ($\%$)&  $\Delta$B$_\parallel$ ($\%$)\\  \hline
B$_0$ & 90 & 92 & 400 & 2090 & 3.997 & 0.00171 & 0.013\\ 
B$_1$ & 30 & 43 & 25 & 1820 & 0.749 & 0.34928 & 6.343\\ 
B$_x$ & 70 & 72 & 300 & 6200 & 3.854 & 0.00287 & 0.074\\
B$_y$ & 120 & 123 & 400 & 6300 & 2.997 & 0.00079 & 0.019\\ 
B$_z$ & 55 & 55 & 200 & 5150 & 3.270 & 0.00038 & 0.121\\ \hline
\end{tabular*}
  \caption{Specification of the \mbox{(pseudo-)Helmholtz coils}. Technical properties: R denotes the coil radius and d the distance of the coil pair. N$_{\text{win}}$ is the number of windings. R$_{\Omega}$ gives the ohmic resistance which consists of the intrinsic resistance of the wire and an additional resistor. Magnetic properties: B gives the magnetic flux for a given current. $\Delta$B$_\bot$ and $\Delta$B$_\parallel$ is the percentage inhomogeneity of the magnetic flux within the cylindrical probe volume ($\varnothing$ = 6~mm, height = 20~mm) induced by one coil pair. $\Delta$B$_\bot$ and $\Delta$B$_\parallel$ are the perpendicular and parallel components with respect to the laser beam.}
  \label{tab:tab01}
\end{table*}

\begin{figure}[!hbt]
\includegraphics[width=\columnwidth]{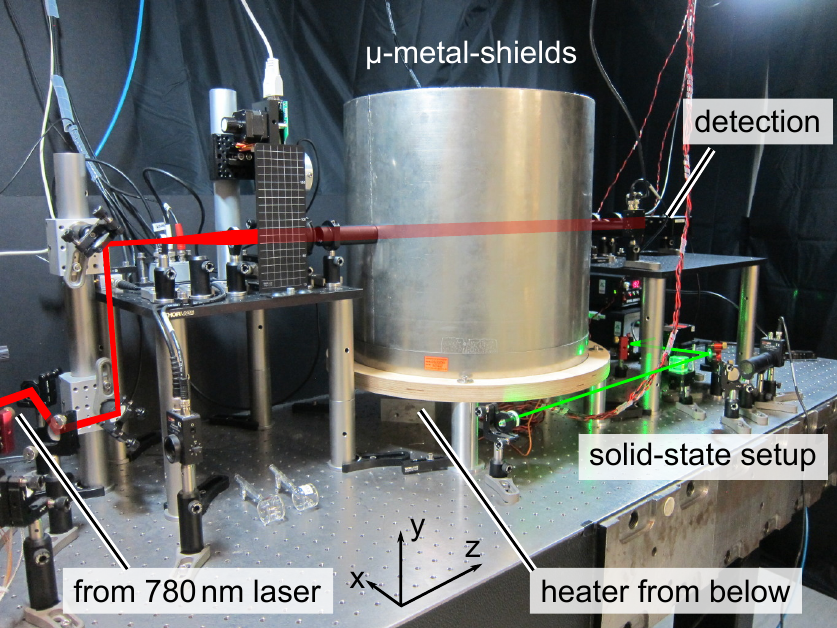}
\caption{Optical setup with labeled optical path of the red laser (red) and four mounted $\upmu$-metal shields. The future solid-state spin setup is indicated at the bottom in green, and is optically connected from below the shielding.}
\label{fig:fig03}       
\end{figure}

\noindent
The inner sensitive part with all coils and the vapor cell is shielded from the environment by four $\upmu$-metal shields (Sekels GmbH, Germany). The inner diameter of the smallest shield is 300~mm. The coils described above are still separated from this shield by at least 30~mm. No further high-frequency suppression, such as Al or Ferrite shielding, is applied. The total shielding setup suppresses external magnetic fields by a factor of 10$^6$, according to the manufacturer.

All shields are vertically mounted on their lids, which are held on the optical table with a wooden support. Five 15~mm holes from the bottom allow for an access from below. These are used for the temperature sensor, the electrical connections to the coils, and the optical and microwave access to the solid-state setup.

The chamber of the inner-most shield is heated by a hot air flow, which is implemented using an SMD-soldering station (Conrad Electronics, Hirschau, Germany) and a supply of compressed air. It is controlled by an external temperature controller (CNPT Series, Omega, Samford, CT, US). The temperature is measured close (50~mm) to the Rb-cell. No influence in the magnetometer signal is observed with and without the temperature sensor. Hot air scintillation is suppressed by 25~mm $\varnothing$ optical borosilicate windows, which are mounted on the most inner shield. In earlier experiments these optical shieldings were not in place and laser power fluctuations were larger. The hot air leaves the shielding via the bottom ports. The temperature of the cells was set to 55\degree C for all measurements below. This corresponds to an atomic density of $2.24\times10^{11}$ atoms per cm$^3$~[\onlinecite{steck_rb85}].

\begin{figure}[!hbt]
\includegraphics[width=\columnwidth]{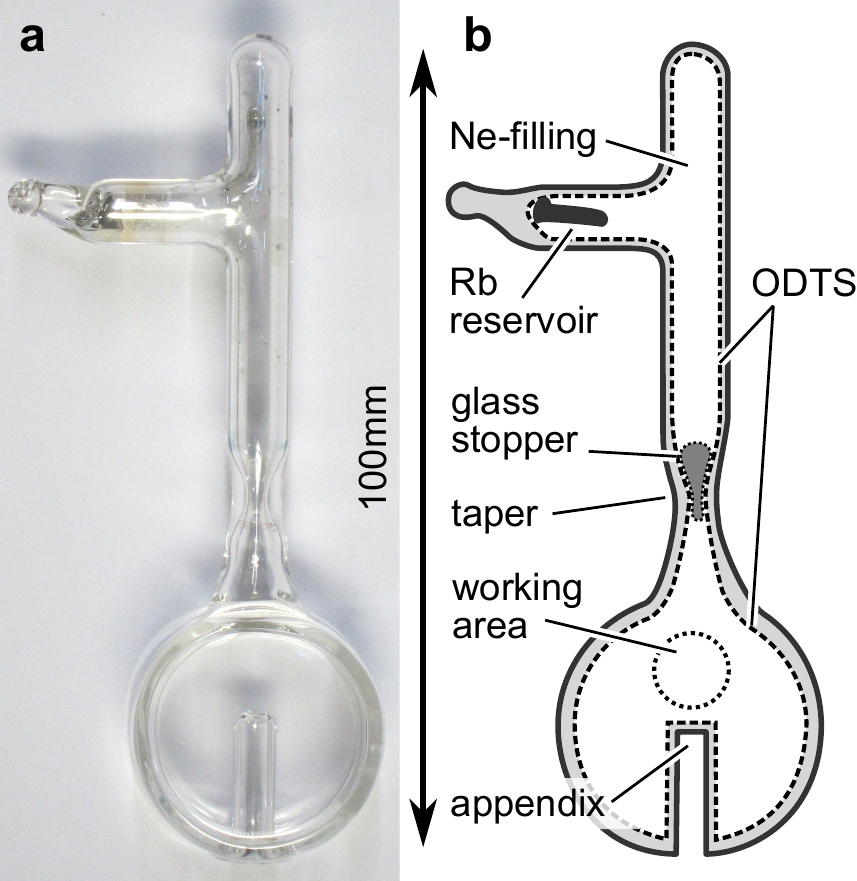}
\caption{Front view of a typical Rb-cell. The appendix for the solid-state sample is located at the bottom of the cell. An additional side arm was added for the possibility for opening the glass stem and heating up the horizontal side arm to transfer some rubidium atoms into the cell volume. A glass stopper limits the diffusion of Rb-atoms to the Rb-reservoir. The cell windows are coated with octadecyltrichlorosilane (ODTS), which is used to suppress collisional spin relaxation. 45~torr Ne-buffer gas is added to suppress atom-atom collisions.}
\label{fig:fig04}       
\end{figure}

Special attention was given to the cell design of the magnetometer. Due to our previous experiences in fabricating atomic vapor cells, we kept our common 26~mm inner diameter design and an optical length of the cells of 20~mm. The wall thickness is 2~mm. One of the produced cells is depicted in Fig.~\ref{fig:fig04}. It was designed such that it is able to house a tiny solid-state sample and allows for some optical, and electrical access. The atomic probe volume should be close, but eventually inhomogeneously distributed around the sample. All this is realized by a small appendix of the Rb-vapor cell. It is made of 500~$\upmu$m thick glass tubing with an inner diameter of slightly larger than 2~mm and a length of 10~mm. Optical shielding of the sample against the magnetometer is realized with opaque carbon fiber material. The cell was designed to allow for a beam-diameter of 10~mm. This gives an atomic probe volume of approx.\ 1600~mm$^3$.

Due to the reduced dimension of the cell and the desired sensitivity of the magnetometer, we introduced some measures to increase the spin-relaxation time in the following:

The Rb-reservoir, hosting some solid rubidium, and the cell are mechanically separated. Since solid rubidium in close proximity to the measurement region reduces the spin coherence time, the two compartments of the cell were heated prior to the experiment with a heat gun such that no solid rubidium is in the measurement cell. Note that this procedure takes a long time (approx.\ 20 hours) because of inhibited diffusion in the cells due to the buffer gas (see below). Afterwards, the cells are homogeneously heated in an oven to the working temperature (approx.\ 55\degree C) for 24 hours. Then, a little glass stopper is mechanically brought into place, so that both compartments are separated~\cite{balabas_prl_2010}.

One way of increasing the spin relaxation life-time is to suppress diffusion in the cell and to reduce unwanted collisions with the cell walls. Light inert atoms, such as helium or neon, are typically used for this task. Their spin relaxation cross section is low~\cite{franzen_pr_1959}. The pressure is calculated by the number of spin relaxing collisions per unit time. For quenching the excited state population, sometimes nitrogen gas is added to the cells. Here we use neon buffer gas since helium tends to diffuse out of the cells. No further nitrogen was added to the cells. The neon pressure for the cell described in this work cell was 45~torr.

Another way of increasing the spin relaxation life-time is coating the walls of the vapor cell with organic compounds such as paraffin~\cite{balabas_prl_2010,corney_book_2006}. Several different schemes have been described in literature. Here we use coating with octadecyltrichlorosilane (ODTS, CAS: 112-04-9)~\cite{seltzer_jap_2008}. The cells were produced by our glass blower's workshop in such a way that the cell was intact and only the cell stem was open. Then carefully cleaned with the RCA (developed by Radio Corporation of America) standard procedure and dried in an oven at 80\degree C. A 0.2\% solution of ODTS in dried toluene was applied for 48~hours at ambient conditions. The cells were then dried in vacuum, and quickly processed by our glass blower workshop. The ODTS coating is reported to allow for up to 2000 bounces of alkali atoms before they get depolarized. Unlike cells with paraffin coating, these cells are more suitable for higher temperatures. The cells used in the measurements presented in the following aged for about one year.

\section{\label{Experimental Characterization}Experimental Characterization}
The experimental characterization of the complete magnetometer is divided into the following parts: First the cells are characterized, independently of the exact magnetometry configuration. Then, the ground-state Hanle-resonance~\cite{hanle_zp_1924} occurring at zero magnetic field is used to compensate residual fields inside our measurement configuration~\cite{castagna_pra_2011}. Afterwards, non-zero magnetic field measurements are demonstrated. Firstly, the magnetic field is changed and the cell's achievable linewidth is measured. Secondly, the non-zero magnetic field is kept constant and the achievable signal-to-noise ratio (SNR) is measured. This all comprises the essentials of the magnetometer sensitivity $\updelta \mathrm{B}$.

\begin{figure}[!hbt]
\includegraphics[width=\columnwidth]{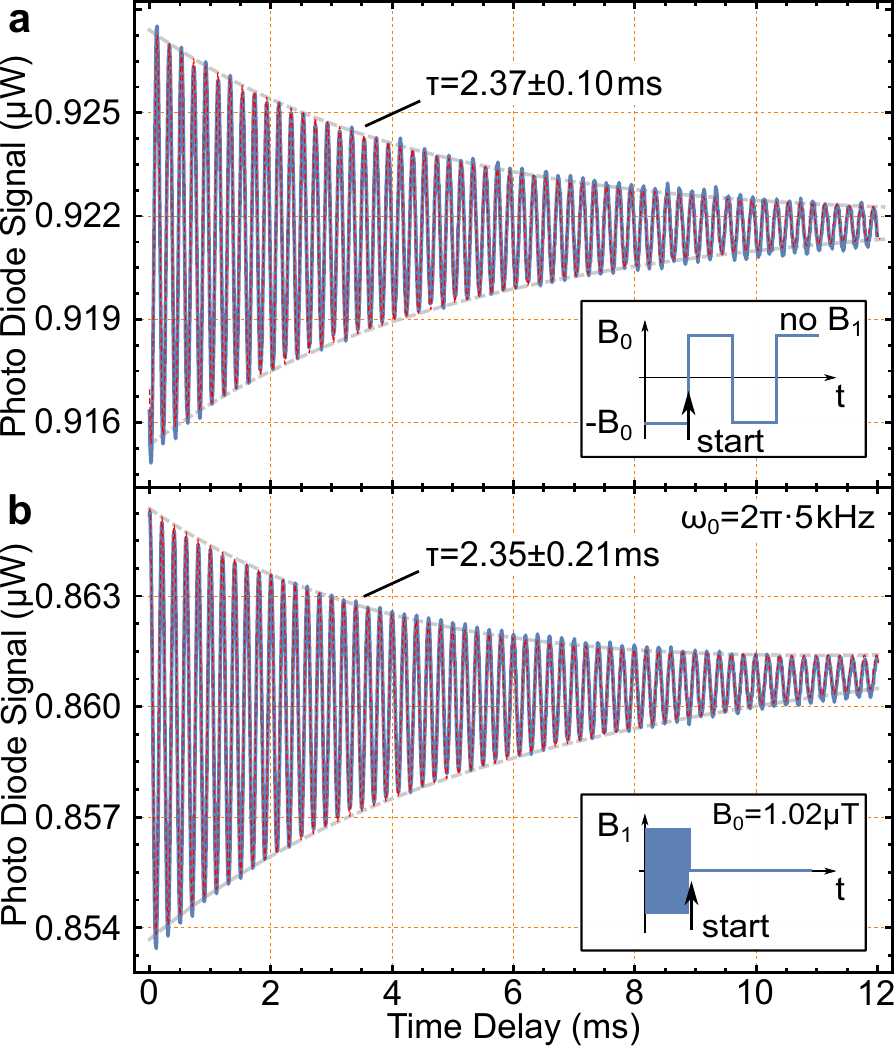}
\caption{Lifetime measurements of the spin coherence time at low laser power of 5~$\upmu$W. (a) The B$_{0}$ field is switched on and off. The atomic system starts after the perturbation with some oscillations and the Larmor frequency is reached. The curve is fitted by Eqn.~\ref{eqn:eqn01} (red dashed line). (b) At a static B$_{0}$-field, corresponding to a Larmor frequency of 5~kHz, the B$_{1}$-field is used to drive the atoms coherently to their Larmor precession. At time zero, the B$_{1}$-field is switched off. Data is fitted by Eqn.~\ref{eqn:eqn02}. The achievable (half-)linewidth (HWHM) of the implemented magnetometer is approx.\ $1/(2 \times 2 \pi \tau)$=34~Hz.}
\label{fig:fig05}       
\end{figure}

The cell design as described above might introduce a reduced spin relaxation time due to the appendix, which is very close to the measurement area, that is illuminated by the Rb-resonant laser beam. There are several ways to measure the spin-relaxation time, which utilize more than one beam and perform a pump-probe measurement~\cite{balabas_prl_2010}. A simplified measurement with one beam allows for two measurement types performed here. These methods are performed within the magnetometer -- but the spin-relaxation time is an intrinsic property of the cell, thus the magnetometer is described in full below.

One way of performing a lifetime measurement is to quickly turn on a magnetic field and observe the system's coherent response. The atoms start to precess in a coherent fashion after the initial ``kick'', but their coherence is lost over time (cf.\ Fig.~\ref{fig:fig05}a) due to spin-relaxation mechanisms like spin-destruction collisions (between alkali- and buffer-gas-atoms and the cell wall), spin-exchange collisions between the alkali-atoms, field gradients across the cell, and interactions with the polarizing light beam~\cite{seltzer_phd_2008}. Usually, the system needs some time to align to the corresponding Larmor frequency. At the end of the relaxation process, the Larmor precession with frequency $\omega_{\mathrm{L}}$ is tuned to the magnitude of the external field. No further oscillation, e.g.\ an external B$_1$-field, apart from the initial switch-on, is applied in this scheme. One crucial point is that the transient time to switch on the magnetic field has to be significantly smaller than the relaxation time. The response can be described as a damped oscillator~\cite{demtroeder3_book_2005} with some initial transience by assuming an exponential frequency approach where the oscillation converges to the correct Larmor frequency $\omega_{\mathrm{L}}$ after a characteristic transient time $\tau_0$: 
The response, a damped oscillation with some initial transient, is described by:

\begin{equation}
\xi(t) = a \sin{(\omega_{\mathrm{L}} (1 - e^{-\frac{t + t_0}{\tau_0}}) t + \varphi)} e^{-\frac{t}{2\tau}} + m t + b
\label{eqn:eqn01}
\end{equation}
$a$ denotes the amplitude of the oscillation at time $t=0$ and $\varphi$ is its phase. The parameter $t_0$ determines the initial frequency. Since a positive slope $m$ is inherent in the signal's envelope, a linear term is added. $b$ is the background offset.
The switching of the B$_0$-field results in an observed characteristic spin-lifetime $\tau=2.37 \pm 0.10$~ms. This corresponds to an expected natural linewidth (HWHM) of 34~Hz.

Another common way is to align an external magnetic field (B$_0$) offset related to a desired Larmor precession and to drive the atoms coherently with a resonant external sinusoidal radio frequency field (rf), that is applied via the B$_1$-coils. After the driving field is switched off, the intrinsic Larmor precession decays for the characteristic spin relaxation time. This is shown in Fig.~\ref{fig:fig05}b. This case is described by a simple damped oscillator without any time dependence of the frequency:

\begin{equation}
\xi(t) = a \sin{(\omega_{\mathrm{L}} t + \varphi)} e^{-\frac{t}{2\tau}} + m t + b
\label{eqn:eqn02}
\end{equation}

The resulting lifetime $\tau$ of 2.35~ms is compatible with the prior measurement. The cells are now characterized inside the magnetometer configuration.

\begin{figure}[!hbt]
\includegraphics[width=\columnwidth]{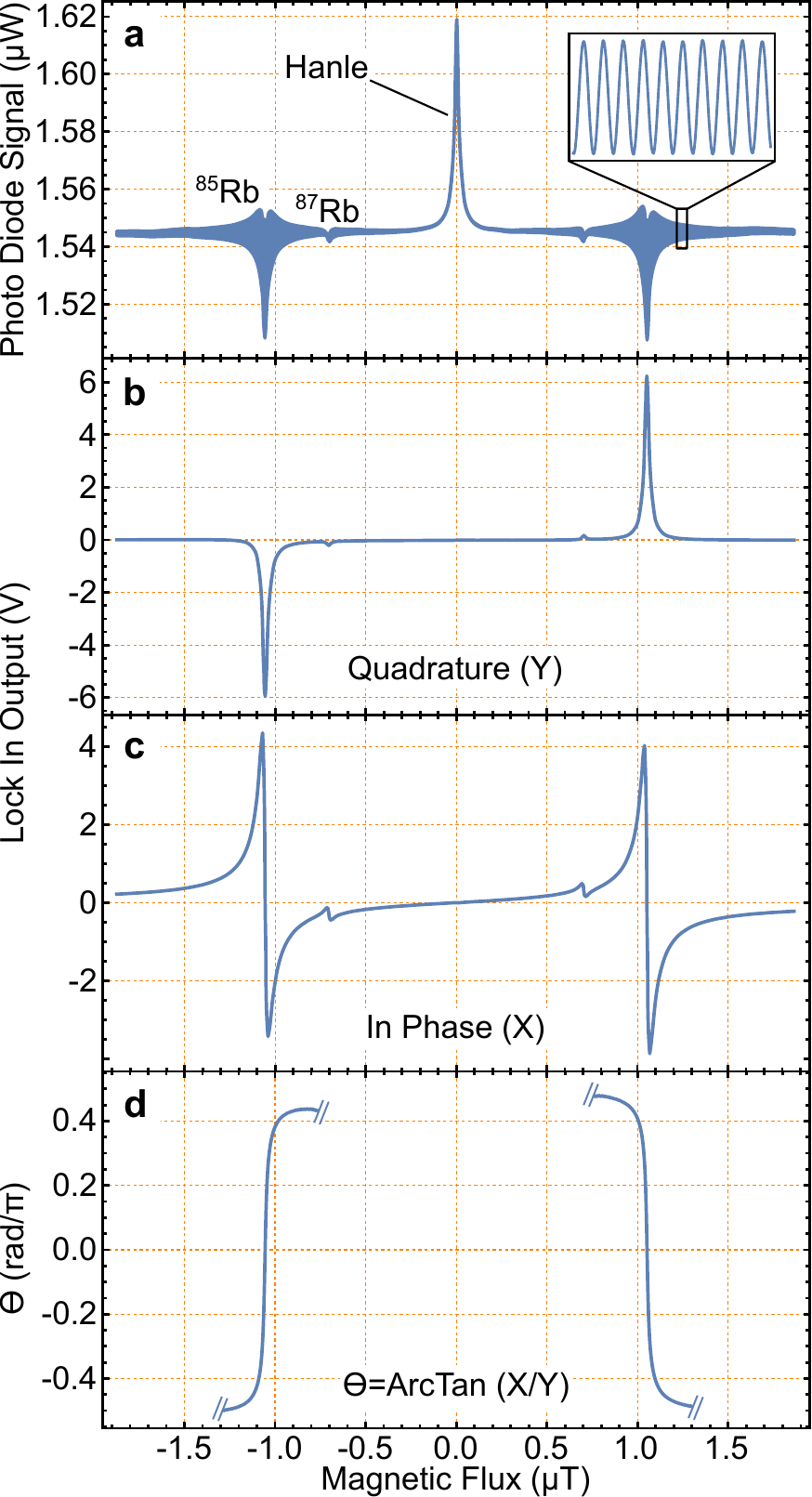}
\caption{Raw data of the Rb magnetometer output for the 45\degree -M$_{\mathrm{x}}$ configuration at a small laser power of 10~$\upmu$W and compensated residual fields. For a fixed B$_{1}$ frequency at 5~kHz, the Rb vapor was exposed to a slow B$_{0}$ sweep executed over a period of 20~sec. (a) shows the Hanle resonance centered around B$_{0}$ = 0 and B$_{1}$ resonances when B$_{0}$-field corresponds to an atomic transition of the B$_{1}$-Larmor-frequency at 5~kHz. The inset is a zoom into the resonance spike for visualization of the 5~kHz-oscillation. (b) and (c) are the lock-in quadrature (Y) and in phase (X) outputs, respectively, for a lock-in time constant of $100$~$\upmu$s and a sensitivity of 20~mV. (d) is the phase signal $\Theta$ obtained from the ratio of both lock-in outputs and Eqn.~\ref{eqn:eqn_lockin_phi}.}
\label{fig:fig06}       
\end{figure}

Although the prior measurements are performed inside our magnetometry configuration, their relevance is independent of the specific implementation of the magnetometer which is introduced now.\\
After inserting the cells into the magnetometer, an important first step is the compensation of residual fields inside the shielding chamber, since every rearrangement is influencing the remanent magnetization. Since we do not have an option to demagnetize the chamber, we rely on the compensation coils (B$_x$, B$_y$, B$_z$) described above. This is commonly performed by monitoring the Hanle resonance~\cite{castagna_pra_2011}, which appears at zero magnetic field. It describes the polarization of the atoms and the orientation of the quantization axis. The height, the sign, and the width are indicators for a correct compensation of the fields. 

The procedure to compensate residual fields (based on the set-up introduced above) is as follows: 
\begin{enumerate}
\item
A longitudinal magnetic field (B$_z$ with respect to the laser beam, $z$-direction) is scanned over zero. Then, the transversal field components (B$_x$, B$_y$) are used to minimize the amplitude of the observed Hanle resonance. In the ideal case, the Hanle resonance vanishes.
\item
A transversal magnetic field is scanned over zero in the $x$- or $y$-direction. The resulting residual field is compensated by adjusting the longitudinal field component in the $z$-direction.
\end{enumerate}

To verify a successful residual field compensation, the (45\degree-)B$_0$ field is scanned spanning over zero magnetic field. Usually, in addition, a magnetic sinusoidal rf B$_1$ field is applied. As calculated by the magnetic Bloch-equations, the Hanle resonance should point upwards and be centered around the zero field crossing while Larmor resonances corresponding to the B$_1$ frequency are observed symmetrically around zero (cf.\ Fig.~\ref{fig:fig06}a). An automated compensation process is easily implemented by software.

One crucial parameter of the magnetometer sensitivity is the magnetic resonance linewidth $\Delta \nu_{\mathrm{HWHM}}$ which can easily be obtained by performing a measurement with phase-sensitive detection using a lock-in amplifier which the photo-diode signal is fed into. To achieve this, a static rf B$_1$ field and a B$_0$ field scan across Larmor resonances are applied to the polarized Rb vapor.

\begin{figure}[!hbt]
\includegraphics[width=\columnwidth]{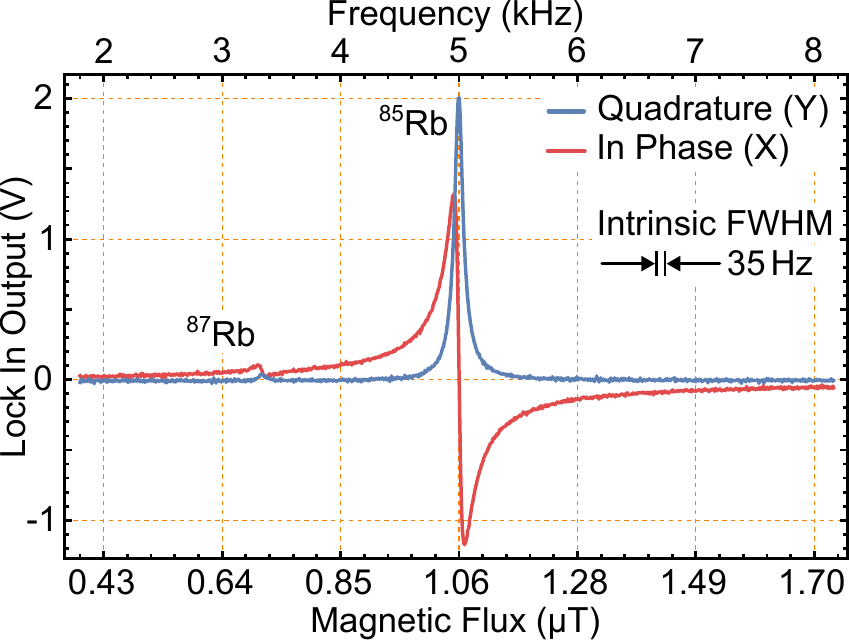}
\caption{In-phase {X} and quadrature {Y} component of the lock-in output for the 45$\degree$ M$_{\mathrm{x}}$-magnetometer at an incoming laser power of 5~$\upmu$W. A small magnetic radio frequency field B$_{1}$ was applied at 5~kHz with root-mean-square (rms) amplitude of 0.012~mA. The B$_{0}$ field was scanned across the resonance from 0 to 1.9~$\upmu$T within 10~sec. A lock-in with time constant of 100~$\upmu$s and 10~mV sensitivity was used.}
\label{fig:fig07}       
\end{figure}

We generate a second sine wave with equal frequency and phase but higher amplitude and use it as a reference signal for the lock-in amplifier. This allows the application of very weak rf B$_1$ fields since the phase-detection via the lock-in is amplitude limited. Fig.~\ref{fig:fig06}b, c, and Fig.~\ref{fig:fig07} show the quadrature output (Y), and the in-phase output (X) of the lock-in amplifier. The width can be extracted by a Lorentzian fit~\cite{groeger_epj_2006} (Eqn.~\ref{eqn:eqn_lockin_Y}) in case of the quadrature, by determining the extrema of the dispersive in-phase output~\cite{groeger_epj_2006} or via a fit of Eqn.~\ref{eqn:eqn_lockin_X}, where $x=(\nu_\mathrm{L}-\nu_\mathrm{rf})/\Delta\nu_{\mathrm{HWHM}}$ is the detuning normalized to the (light-power dependent) linewidth $\Delta\nu_{\mathrm{HWHM}}$ of the resonance relating to the rf magnetic B$_1$ field.
\begin{align}
Y(x) &=-A \frac{x}{x^2+1+S}  \label{eqn:eqn_lockin_Y} \\
X(x) &=-A \frac{1}{x^2+1+S}  \label{eqn:eqn_lockin_X} \\
\Theta(x)&= \arctan{x}  \label{eqn:eqn_lockin_phi}
\end{align}
The broadening of the line due to the power (rms) of the rf field is described by the saturation parameter $S$. The phase signal $\Theta (x)$, which is determined by the ratio of both lock-in outputs $X(x)/Y(x)$, excludes $S$. Hence, the width of $\Theta (x)$ in Eqn.~\ref{eqn:eqn_lockin_phi} is immune to rf power broadening~\cite{groeger_epj_2006}. The frequency can be expressed in the magnetic flux $B$ (and vice versa) with $\nu_\mathrm{L}=(\gamma/2\pi)\times B$. The rubidium 85's gyromagnetic ratio is $\gamma$ = $4.7~\mathrm{Hz/nT}$~[\onlinecite{steck_rb85}].

\begin{figure}[!hbt]
\includegraphics[width=\columnwidth]{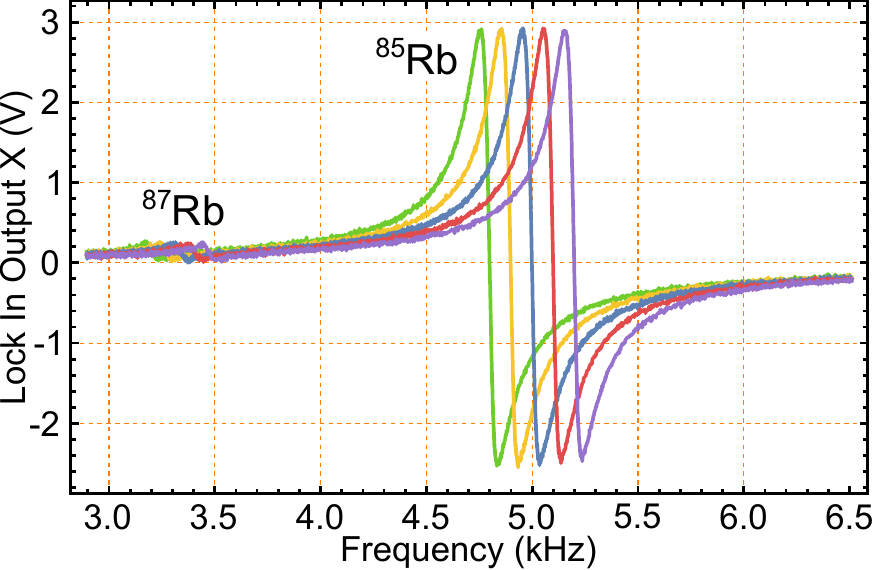}
\caption{In-phase lock-in output (X) for a B$_{0}$ scan at different fixed B$_{1}$ frequencies (4.8-5.2~kHz) with step size of 100~Hz. Same parameters as Fig.~\ref{fig:fig07}. Only the right hand side of the the B$_{0}$ sweep is shown. The opposite side is essentially equal.}
\label{fig:fig08}       
\end{figure}

The spin coherence lifetime and the measured linewidth above provide a first insight into the achievable sensitivity. A change of the Larmor frequency in proximity to the described frequency of $\omega_0$=$2 \pi \times$5~kHz does not degrade or change the signal and the derived linewidth. A quick check of the operation at different frequencies allows for a quick  visualizing of the linewidth, since it can be read off the graph directly. A magnetometer operation on different Larmor frequencies in 100~Hz steps is presented in Fig.~\ref{fig:fig08}. 

\begin{figure}[!hbt]
\includegraphics[width=\columnwidth]{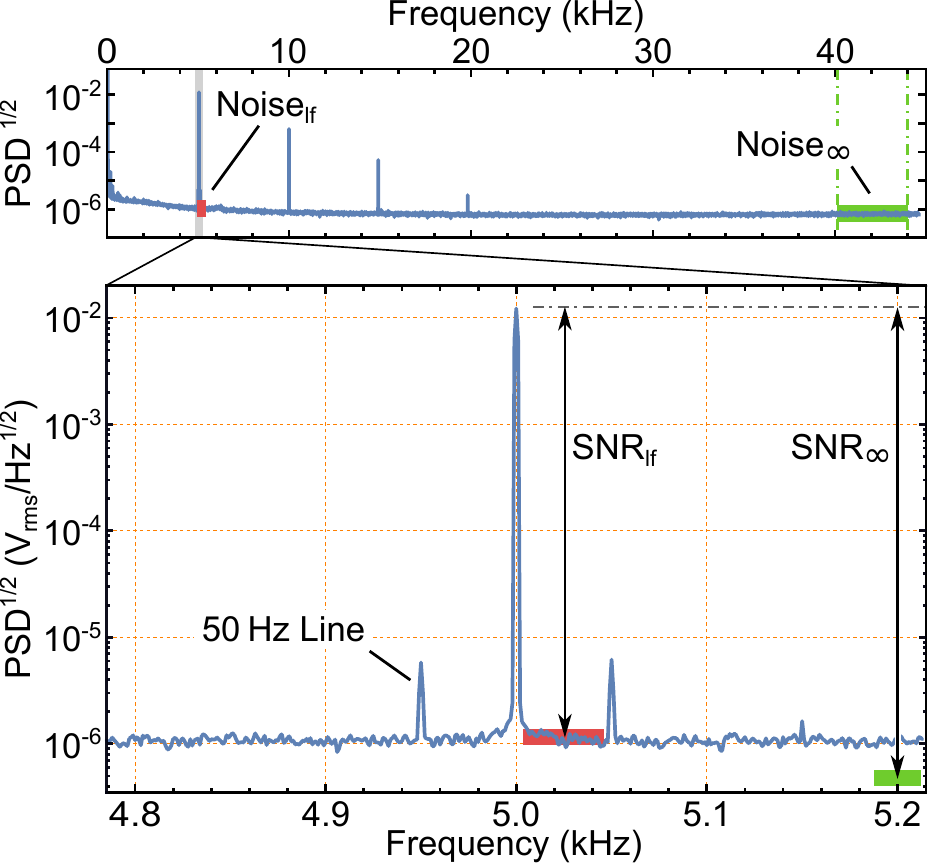}
\caption{Square root of the power spectral density (PSD$^{1/2}$) in free running mode. The average of 45 spectral densities convoluted with a Hanning window function is shown, each resulting from a measurement with 1~sec integration time, thus presenting the noise level for a 1~sec measurement time. The measurement was performed at an incoming laser power of 100~$\upmu$W, a B$_1$ rms amplitude of 0.58~mA and a frequency of 5~kHz with correspondingly resonant static B$_0$ field. Typical 50 and 150~Hz side-peaks are due to noise at the power line frequency. Technical low frequency noise, possibly introduced by laser power fluctuations, imperfectly polarized light, and magnetic field fluctuations~\cite{budker_book_2013} increases the noise around the peak. The average low frequency noise is represented by the red bar. The signal-to-noise ratio SNR$_{\mathrm{lf}}$ refers to this noise level. Considering the decreased noise level at higher frequencies yields the SNR$_{\infty}$. We estimated the high-frequency noise by averaging over the frequency range between 41 -- 44~kHz (green bar).}
\label{fig:fig09}       
\end{figure}

The other important parameter for the sensitivity is the signal-to-noise ratio (SNR) of the acquired signal. To estimate the SNR, a measurement is performed in which the alkali atoms are driven coherently in resonance with the Larmor frequency $\omega_\mathrm{L}$, which is set by the static magnetic field B$_0$. The photo-diode signal is acquired over a time span of 1~sec. Therefore, the unit $\sqrt{\mathrm{Hz}}$ of the SNR is omitted in the following. The spectral density (Fig.~\ref{fig:fig09}) is calculated using a Hanning window. The power spectral density was integrated over the width of 3~Hz of the signal-peak (here at 5~kHz). The root mean square (rms) value times $\sqrt{2}$ gives the amplitude in Volt (V). Since the high frequency noise levels (Noise$_{\mathrm{\infty}}$) are significantly smaller than the low frequency noise levels (Noise$_{\mathrm{lf}}$), we have estimated two signal-to-noise ratios. For the SNR$_{\mathrm{lf}}$ we chose the Noise$_{\mathrm{lf}}$ in immediate proximity to the signal-peak by averaging the noise floor between the signal-peak and the 50~Hz side-peak of the power line from 4960 to 4995~Hz, as indicated in Fig.~\ref{fig:fig09}. This noise level consist of low-frequency components caused by magnetic field fluctuations around the signal-peak.
For SNR$_{\mathrm{\infty}}$ the noise Noise$_{\mathrm{\infty}}$ is used, corresponding to the average noise level between 41 and 44~kHz.


The full sensitivity of the magnetometer is a combination of the experimental values derived above. It is calculated as

\begin{equation}
\delta \mathrm{B}=\frac{\Delta \nu_{\mathrm{HWHM}}}{\gamma \cdot \mathrm{SNR}} \vspace*{1mm}
\label{eqn:sensitivity}
\end{equation}

\noindent with the gyromagnetic ratio $\gamma$.

With the aforementioned described measurements, the sensitivity for our setup is calculated to $\delta \mathrm{B}_{\infty}$ = 0.6~pT/$\sqrt{\mathrm{Hz}}$ at an incoming laser power of 0.125~mW. $\delta \mathrm{B}_{\infty}$ is determined using the SNR$_\infty$ of 39700, a width $\Delta \nu_{\mathrm{HWHM}}$ of 111~Hz, and $\gamma = 4.7~\mathrm{Hz/nT}$.
A second sensitivity $\delta B_{\mathrm{lf}}$ is determined referring to the SNR$_{\mathrm{lf}}$. For this sensitivity, our optimum $\delta B_{\mathrm{lf}}$ of 1.5~pT/$\sqrt{\mathrm{Hz}}$ was found at an incoming laser power of 0.05~mW, resulting in an SNR$_{\mathrm{lf}}$ of 12800 and a width $\Delta \nu_{\mathrm{HWHM}}$ of 88~Hz. These values are derived from the optimization process which is described in the following.

\section{\label{Optimization}Optimization}

As described above, the sensitivity is determined by an interplay between the magnetic resonance HWHM and the SNR. The best magnetometer sensitivity is obtained at the smallest ratio between them. A variety of effects influence these parameters. Examples are the number of polarized (pumped) atoms, the coherence of the Larmor precession between the single atoms, the influence of the rf B$_1$ field, magnetic field-inhomogeneities, noise etc.. All these (internal) influences are summarized in the linewidth and the signal-to-noise ratio. In the following, we present an optimization routine to achieve the highest sensitivity of the magnetometer configuration. Although there are more parameters which in principle can be taken into account (beam-diameter, density of the vapor). A change of the most evident parameters is presented. The best sensitivity dependent on the power of the rf B$_1$ field, and the incoming laser power is determined below.

\begin{figure}[!hbt]
\includegraphics[width=\columnwidth]{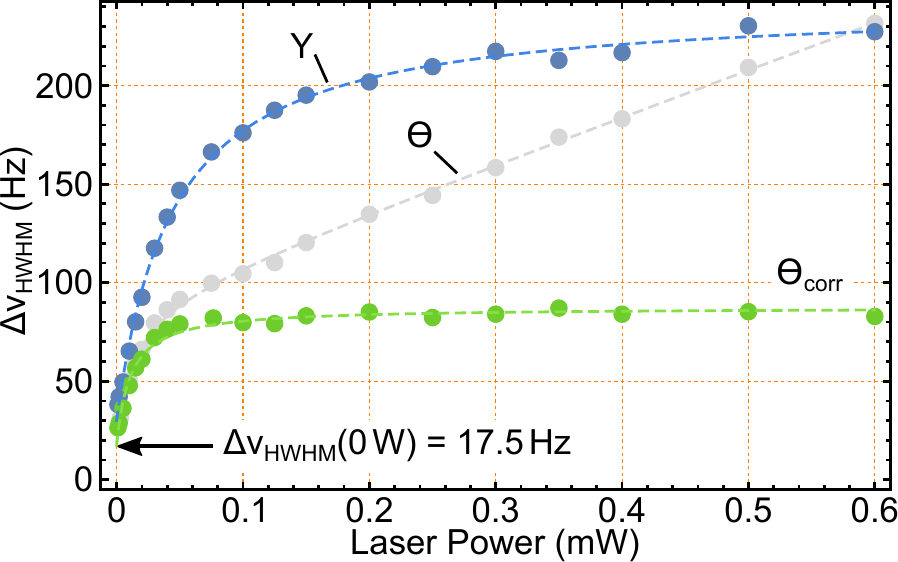}
\caption{Cell's linewidth (HWHM) over input laser power. The blue dots represent the linewidth of the quadrature output (Y) of the lock-in estimated by a Lorentzian fit. Fitting by Eqn.~\ref{eqn:hwvspower} yields the blue dashed line. The gray dots result from a cyclometric function fit to the phase signal $\Theta$ (cf.\ Fig.~\ref{fig:fig06}). Even though $\Theta$ excludes B$_1$ power broadening, we still see an artificial linear broadening which might be originated in an increasing technical noise, magnetic field fluctuations or light power broadening. The dashed gray line is a fit using Eqn.~\ref{eqn:hwvspower} with an additional linear term. Subtracting the latter yields a corrected dataset $\Theta_{\mathrm{corr}}$ without linear broadening effects (green dots). The black arrow indicates the intrinsic linewidth at zero power, extrapolated over the fit of $\Theta_{\mathrm{corr}}$.}
\label{fig:fig10}       
\end{figure}

The linewidth $\Delta \nu_{\text{HWHM}}$ of the phase signal $\Theta$, which has been introduced above, is immune to rf B$_1$ power broadening for a wide range. Hence, only the laser power dependence of $\Delta \nu_{\text{HWHM}}$ is measured. The data is shown in Fig.~\ref{fig:fig10}. The data points of the phase $\Theta$ are fitted by Eqn.~\ref{eqn:hwvspower} and an additional linear term~\cite{appelt_pra_1999}. 
\begin{equation}
\Delta \nu_{\text{HWHM}}(P_\mathrm{l}) = \frac{ \nu_\mathrm{sat} \cdot P_\mathrm{l}}{(P_{\mathrm{sat}} + P_\mathrm{l})} + \nu_0
\label{eqn:hwvspower}
\end{equation}
$P_\mathrm{l}$ denotes the incoming laser power, $P_\mathrm{sat}$ is a saturation power and $\nu_0$ the width at zero power (or the cell's intrinsic linewidth), and $\nu_\mathrm{sat}$ the saturation width.

\begin{figure}[!hbt]
\includegraphics[width=\columnwidth]{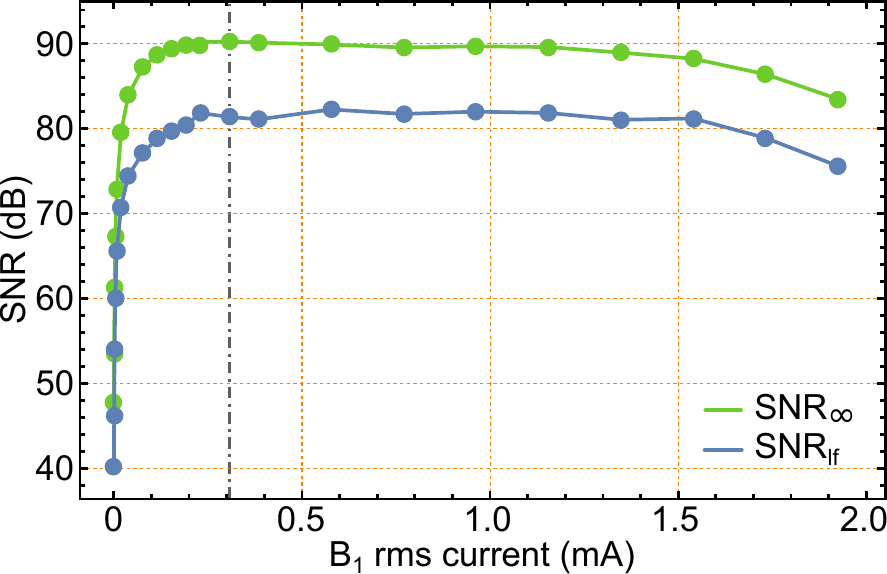}
\caption{Dependence of the SNR on the rms current of the rf B$_1$-coil with a fixed frequency equivalent to the Larmor frequency $\omega_\mathrm{L}$ of 5~kHz, which is set by a static magnetic field (B$_0$). The SNR$_{\infty}$ (green line) is estimated by the noise average from 41 to 44~kHz. The blue line shows the SNR$_{\mathrm{lf}}$ with a noise average from 5 to 40~Hz away from the carrier (cf.\ Fig.~\ref{fig:fig09}). The gray line indicates the best SNR with a B$_1$ rms current of $0.31~\mathrm{mA}$. The measurement was performed at an incoming laser power of $100~\mathrm{\upmu W}$. }
\label{fig:fig11}       
\end{figure}

As a second step, a measurement over the power of the rf B$_1$ field versus the related SNRs is performed at an arbitrary constant laser power (cf.\ Fig.~\ref{fig:fig11}). The power of the B$_1$ field is changed until a clear maximum SNR appears. The SNRs are estimated from the PSD (cf.\ Fig.~\ref{fig:fig09}) as mentioned above. Based on these measurements, the rf B$_1$ field power of the best SNR is taken to perform a further measurement sequence in which the power dependence of the SNRs is evaluated.

\begin{figure}[!hbt]
\includegraphics[width=\columnwidth]{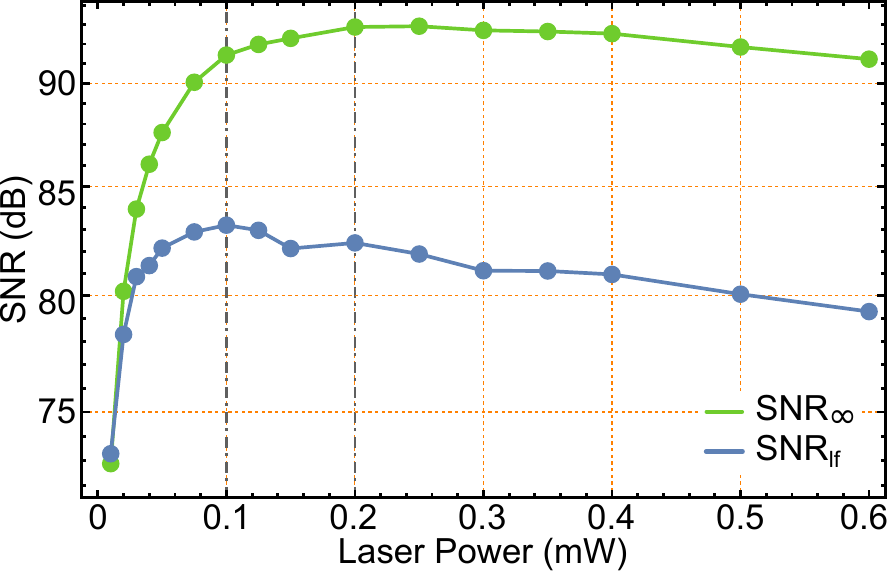}
\caption{Dependence of the SNR on the incoming laser power with a B$_1$ coil rms current of 0.31~mA, corresponding to the measurement shown in Fig.~\ref{fig:fig11}. The B$_1$ frequency was set to 5~kHz, in accordance to the Larmor frequency induced by the static B$_0$-field. The maxima are located at laser powers of $0.1~\mathrm{mW}$ and $0.2~\mathrm{mW}$, respectively, as indicated by the gray dashed lines.}
\label{fig:fig12}       
\end{figure}

The same measurements are performed as before, with the exception of changing laser power while keeping the power of the rf B$_1$ field constant. It should be noted, that the gain settings of the photo diode were not changed from 30~dB in the course of the measurement, since different gain settings imply a different noise level. The result of the measurements is shown in Fig.~\ref{fig:fig12}.

\begin{figure}[!hbt]
\includegraphics[width=\columnwidth]{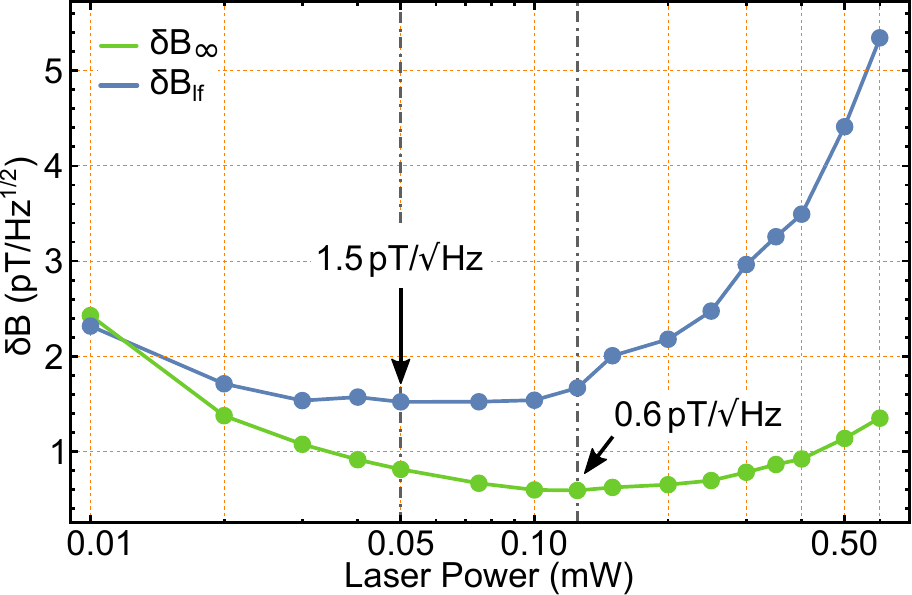}
\caption{Power dependence of the magnetometer's sensitivity $\updelta \mathrm{B}$ determined by Eqn.~\ref{eqn:sensitivity} and the data shown in Fig.~\ref{fig:fig10} and Fig.~\ref{fig:fig12}. We used the width of the phase signals $\Theta$ from Fig.~\ref{fig:fig10} since it excludes B$_1$ power broadening. $\updelta \mathrm{B}_{\infty}$ (green line) and  $\updelta \mathrm{B}_{\mathrm{lf}}$ correspond to the signal-to-noise ratios shown in Fig.~\ref{fig:fig12} with their best sensitivities of 1.5 and 0.6~$\mathrm{pT}/\sqrt{\mathrm{Hz}}$ at 0.05 and 0.125~mW laser power, respectively.}
\label{fig:fig13}       
\end{figure}

The both measurement sequences over the laser power of the $\Delta \nu_{\text{HWHM}}$ (Fig.~\ref{fig:fig10}) and the SNR (Fig.~\ref{fig:fig12}) allows a determination the related sensitivities with Eqn.~\ref{eqn:sensitivity}. The obtained curves in Fig.~\ref{fig:fig13} exhibit minima for the best sensitivity.  
It should be mentioned that in case of a sufficient time capacity, a more accurate optimization can be performed by a two-dimensional-series approach with the B$_1$-amplitude, and the laser power. 

\section{\label{ConclusionOutlook}Conclusion \& Outlook}
In conclusion, we have presented and characterized a compact and simple M$_{\mathrm{x}}$-magnetometer based on atomic rubidium. A focus was given to a special cell design, which features an anti-relaxation coating, and neon buffer gas. In terms of magnetometer sensitivity, the cells fail to perform at the level compared to when they were first produced one year ago. Additionally, they are 10-100~$\times$ less sensitive to comparable cell describes in literature~\cite{seltzer_jap_2009}. On the other hand, typically cell designs are larger and do not incooperate any collision enhancing glass features, such as the appendix, which can host a solid state sample.

The magnetometer's sensitivity is determined to be 0.6~pT/$\sqrt{\mathrm{Hz}}$. This is based on an independent estimation of the Larmor spin precession linewidth and the achievable signal-to-noise ratio in our experimental configuration. Since the SNR was determined in a free-running magnetometer configuration, we believe that it will be possible to suppress the low-frequency noise components even more. The magnetometer's performance is characterized in slow B$_0$-scans and in 1~sec time-series recordings. With regard to the experimental equipment the photo-diode configuration for the measurement is not ideal, and we experienced different noise levels for different amplification ratios. A photo-diode design based on a sensitive transimpedance amplifier (DLPCA-200, Femto GmbH) will be implemented in the future. Another important improvement is the in-house design of a low-noise current source, which targets the 10$^{-7}$ relative accuracy regime with a maximum range of a few mA. For the compensation of field gradients, we will divide the \mbox{(pseudo-)Helmholtz} coil pairs into independent coils, which are then independently driven by current drivers.

The magnetometer is prepared to be combined with solid state spin samples. A small appendix ($\varnothing$ 2~mm, 10~mm depth) with sub-mm thick walls can host another experiment. At present, we have prepared an optically shielded, fiber-coupled nitrogen-vacancy ensemble with a microwave loop antenna, which can be inserted into the appendix. Earlier designs featuring a hole through the entire glass cell, were not pursued further, but may eventually allow for the application with liquid samples~\cite{ledbetter_jmr_2009}. For these combination experiments, the implementation of a light-driven Larmor precession (Bell-Bloom type~\cite{bell_prl_1961}) will allow for exciting the atomic spin ensemble completely independent of the solid-state sample.

\begin{acknowledgments}
The authors thank the atomic physics group in Fribourg, Switzerland for their strong support and transfer of knowledge. To name is the support not only from A.\ Weis, but also from V.\ Lebedev, and Z.\ Grujic. Special thanks go to R.\ Jim\'enez Martinez from the Mitchell group at the ICFO in Barcelona, Spain for sharing his knowledge and hands on experience. M.\ Mitchell is acknowledged for his hospitality in his group. Helmut Kammerlander is strongly acknowledged for the production of the high quality glass cells used in this study. We acknowledge the work performed by S.\ Lasse. We further acknowledge the funding from the MPG via a Max Planck fellowship (J.W.), the SFB project CO.\-CO.\-MAT/TR21, the Bundesministerium f\"ur Bildung und Forschung (BMBF), the project Q.COM, and SQUTEC. Lastly, D.\ Arnold enjoyed support from the COST action MP1403 -- Nanoscale Quantum Optics.
\end{acknowledgments}


%
\end{document}